\documentclass[aps,prl,reprint]{revtex4-1}

\usepackage{graphicx}
\usepackage{dcolumn}
\usepackage{bm}

\usepackage[utf8]{inputenc}
\usepackage[ngerman,english]{babel}
\usepackage{lmodern,hfoldsty,charter}
\usepackage{color}
\usepackage{amsmath}
\usepackage{amssymb}

\usepackage{pdfpages}

\makeatletter
\AtBeginDocument{\let\LS@rot\@undefined}
\makeatother

\begin{document}
 \title{Optimal design strategy for non-Abelian geometric phases using Abelian gauge fields based on quantum metric}
 \date{\today}

 \author{Mark Kremer}
 \thanks{both authors contributed equally.}
 \author{Lucas Teuber}
 \thanks{both authors contributed equally.}
 \author{Alexander Szameit}
 \email{alexander.szameit@uni-rostock.de}
 \author{Stefan Scheel}
 \affiliation{Institut f\"ur Physik, Universit\"at Rostock, Albert-Einstein-Stra{\ss}e 23, D-18059 Rostock, Germany}

\begin{abstract}
Geometric phases, which are ubiquitous in quantum mechanics, are commonly more than only scalar quantities.
Indeed, often they are matrix-valued objects that are connected with non-Abelian geometries.
Here we show how genera\-lized, non-Abelian geometric phases can be realized using electromagnetic waves travelling through coupled photonic waveguide structures.
The wave\-guides implement an effective Hamiltonian possessing a degenerate dark subspace, in which an adiabatic evolution can occur.
The associated quantum metric induces the notion of a geodesic that defines the optimal adiabatic evolution.
We exemplify the non-Abelian evolution of an Abelian gauge field by a Wilson loop.
\end{abstract}

\maketitle

\section{Introduction}
Geometry and quantum mechanics are inextricably linked.
Whenever a quantum system evolves in Hilbert space, its wavefunction acquires a phase that solely depends on the path the quantum system has taken.
This concept of geometric phases was introduced by Sir Michael Berry \cite{Berry} who pointed out that there are instances in which these emerging phase factors cannot be removed by some gauge transformation.
A famous example is the Aharonov--Bohm effect \cite{Aharonov} in which the wavefunction of a charged particle travelling in a loop around a solenoidal magnetic field acquires a phase proportional to the flux through the surface enclosed by the loop, i.e. a line integral over the (Abelian) vector potential.
Here the emerging (Abelian) phase is merely a complex number.

However, any quantum system with degenerate energy levels possesses a far
richer structure as matrix-valued geometric phases \cite{Wilczek}, known as
non-Abelian holonomies. These can occur as soon as the emerging phase depends
on the order of consecutive paths. In contrast to the vector potential in the
Aharonov--Bohm scenario, a non-Abelian gauge field is responsible for the
appearance of these matrix-valued phases.
Such phases are crucial for topological quantum computation
\cite{Zanardi,Pachos}, non-Abelian anyon statistics \cite{Nayak}, and the
quantum simulation of Yang--Mills theories.
Non-Abelian synthetic gauge fields are usually realized in systems where
coupled energy levels naturally lead to the required degeneracy such as in cold
atomic samples \cite{Dalibard} and artifical atoms in superconducting circuits
\cite{Abdumalikov}.
In the case of electromagnetic fields, due to their intrinsic Abelian nature
the required degeneracy is more intricate to design.
One succesful scheme utilized the spin-orbit coupling of polarized light
in asymmetric microcavities \cite{Ma}.

In our work, we focus on a different degree of freedom and synthesize a
non-Abelian geometric phase by implementing adiabatic population transfer
\cite{Unanyan} of light. We employ an integrated photonic structure possessing
dark states, in which a non-Abelian geometric phase associated with a U(2) group
transformation is realized. The quantum metric spanned by the subspace of the
dark states induces a geodesic that defines the optimal adiabatic evolution of
these non-Abelian phases.

\section{Theory}
Gauge fields naturally arise in the context of field theories when demanding the invariance of a field under some transformation.
Invariance under multiplication by a scalar phase factor leads to the concept of Abelian gauge fields, such as the four-vector potential of electromagnetism, with commuting components.
In contrast, matrix-valued phases entail non-Abelian gauge fields where the commutator of the individual components involves the structure constants of the underlying Lie algebra.
A famous example are Yang--Mills theories of particle physics \cite{Cheng}.

A wavepacket evolving in the presence of a gauge field acquires a geometric phase.
For Abelian gauge fields, this is the famous Berry--Pancharatnam phase \cite{Berry,Pancharatnam}.
The non-Abelian generalization is known as the Wilson loop
\begin{equation}
\label{eq:wilson}
W_{\mathcal{C}}=\mathrm{Tr} \left[ \mathcal{P} \mathrm{exp} \left( -\oint_{\mathcal{C}} \mathbf{A}_{\nu} \mathrm{d}x^{\nu} \right) \right],
\end{equation}
which is the trace of the path-ordered ($\mathcal{P}$) exponential of the non-Abelian gauge field $\mathbf{A}_{\nu}$ \cite{Wilson}.
Our system is a non-trivial collection of interacting Abelian subsystems
and, thus, non-Abelian, being
characterized by a Wilson loop $W_{\mathcal{C}} < 2$~\cite{Goldman,GoldmanSpielman}.

In order to implement this concept, one seemingly requires a non-Abelian gauge field.
However, this is not necessary, as a non-Abelian structure naturally appears
whenever the evolution of a quantum system is confined to a degenerate subspace
of some Hamiltonian \cite{Wilczek}.
As a consequence, generating non-Abelian geometric phases is not connected to
the presence of a non-Abelian gauge fields, but to the existence of a
degenerate subspace.


\begin{figure}[h!]
 \center
 \includegraphics[width=8cm]{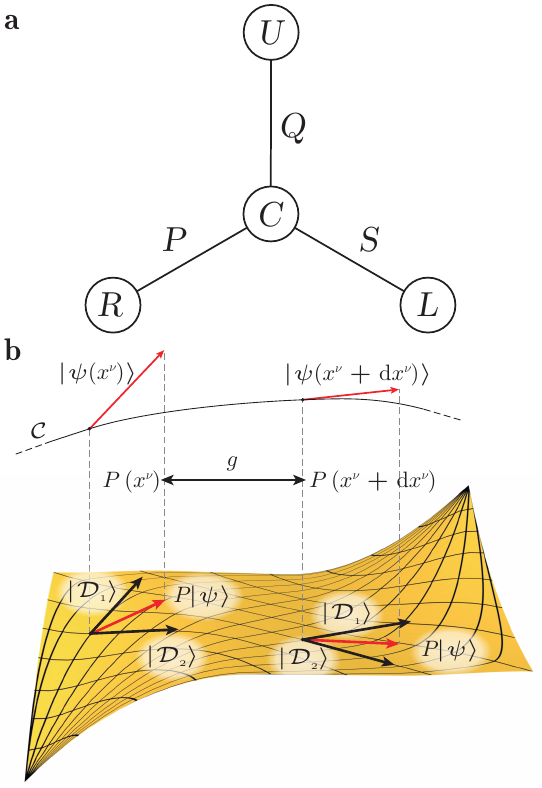}
 \caption{\textbf{a} Scheme of the four-waveguide system.
 \textbf{b} Visualization of changing dark subspace projector along curve $\mathcal{C}$.
 The change of the dark subspace projectors of the system is given by the quantum metric \textbf{$g$}.}
\end{figure}

In our work, we consider the system sketched in Fig.~1 that consists of four potential wells $C, U, R, L$ that are coupled  with time-dependent hopping constants $Q, S, P$.
The Schr\"odinger equation for the field amplitudes $a_{\textrm{\footnotesize{L}}}, a_{\textrm{\footnotesize{U}}}, a_{\textrm{\footnotesize{R}}}, a_{\textrm{\footnotesize{C}}}$ in the individual wells reads, therefore,
\begin{equation}
\label{eq:Hamiltonian}
\mathrm{i} \, \partial_t \left( {\begin{array}{c} a_{\textrm{\footnotesize{L}}}\\a_{\textrm{\footnotesize{R}}}\\a_{\textrm{\footnotesize{U}}}\\a_{\textrm{\footnotesize{C}}}\\ \end{array}} \right)
=
\left( {\begin{array}{cccc}0 & 0 & 0 & S\\0 & 0 & 0 & P\\0 & 0 & 0 & Q\\S & P & Q & 0\\ \end{array}} \right)
\cdot
\left( {\begin{array}{c} a_{\textrm{\footnotesize{L}}}\\a_{\textrm{\footnotesize{R}}}\\a_{\textrm{\footnotesize{U}}}\\a_{\textrm{\footnotesize{C}}}\\ \end{array}} \right)
\end{equation}

This Hamiltonian supports two dark states with zero eigenvalue:
\begin{align}
\label{eq:darkstate1}
|\mathcal{D}_1\rangle &= \sin \theta |w_{\textrm{\footnotesize{L}}}\rangle - \cos \theta |w_{\textrm{\footnotesize{R}}}\rangle, \\
|\mathcal{D}_2\rangle &= \cos \theta \sin \phi |w_{\textrm{\footnotesize{L}}}\rangle + \sin \theta \sin \phi |w_{\textrm{\footnotesize{R}}}\rangle - \cos \phi |w_{\textrm{\footnotesize{U}}}\rangle,
\end{align}
where $|w_{\textrm{\footnotesize{U,R,L}}}\rangle$ are the eigenmodes of the potential wells  $U, R, L$ respectively, with the angle parameterization $\theta=\arctan(P/S)$ and $\phi= \arctan(Q/\sqrt{S^2 + P^2})$.
Notably, they do not involve the eigenstate $|w_{\textrm{\footnotesize{C}}}\rangle$ to which all other states are coupled.
These dark states span a (dark) subspace in which the adiabatic evolution of a wavefunction along a closed path can be described by a non-Abelian geometric phase (\ref{eq:wilson}) with the gauge field
\begin{equation}
\label{eq:gaugeField}
(\mathbf{A}_{\nu})_{ki} = \langle \mathcal{D}_k | \frac{\partial}{\partial x^{\nu}} | \mathcal{D}_i \rangle,
\end{equation}
written in the coordinates $\{x^{\nu}\}=\{S,P,Q\}$ (for details, see Supplementary Materials).
In the context of adiabatic evolution, the Hamiltonian (\ref{eq:Hamiltonian}) is a generalization \cite{Unanyan} of the STIRAP (\textbf{STI}mulated \textbf{R}aman \textbf{A}diabatic \textbf{P}assage) protocol \cite{STIRAP}.
What is required is to ensure adiabaticity of the evolution.

Interestingly, adiabatic transport is equivalent to parallel transport in a curved (metric) space via vanishing covariant derivative \cite{Liu}, i.e. along a geodesic defined in our parameter manifold.
The quantum metric $g_{\mu \nu} = \mathrm{Tr} \left( \partial_{\mu} P \partial_{\nu} P \right)$ is constructed from infinitesimal changes of the dark subspace projector $P(x^\nu)=\sum_i | \mathcal{D}_i (x^\nu) \rangle \langle \mathcal{D}_i (x^\nu) |$ (see Fig.~1~B).
This is the real part of a quantity known as the quantum geometric tensor
\cite{Provost,Rezakhani,Nakahara,Liu}, whose imaginary part is the field
strength tensor of the (non-Abelian) gauge field, $\mathbf{F}_{\mu\nu} =
\partial_{\mu} \mathbf{A}_{\nu} - \partial_{\nu} \mathbf{A}_{\mu} - [
\mathbf{A}_{\mu} , \mathbf{A}_{\nu} ]$.

The coordinates $x^\nu(z)$ in parameter space themselves are a function of the propagation distance $z$.
The quantum metric $g_{\mu \nu}$ defines a path length (action) along the curve $\mathcal{C}$ in the parameter manifold from the input facet $z_{\textrm{\footnotesize{i}}}$ to the output facet $z_{\textrm{\footnotesize{f}}}$,
\begin{equation}
\label{eq:Ldef}
L = \int_{z_{\textrm{\footnotesize{i}}}}^{z_{\textrm{\footnotesize{f}}}} \sqrt{g_{\mu \nu} \partial_z x^{\mu} \partial_z x^{\nu}} \mathrm{d} z.
\end{equation}
The principle of least action defines a geodesic that describes the evolution with the least diabatic error through parameter space \cite{Rezakhani}.
As a consequence, the notion of adiabaticity is intimately connected to the concept of the quantum metric.
This defines the optimal strategy for determining the time dependence of the parameters for adiabatic evolution in parameter space.
Starting from the desire to realize non-Abelian geometric phases, one first has to find a Hamiltonian with a degenerate subspace \cite{Wilczek} on which a metric can be defined.
The geodesic induced by this metric then specifies the variation of the parameters of the Hamiltonian such that the evolution through parameter space occurs with the least diabatic error.
A closed path along the geodesic in parameter space then necessarily results in a non-Abelian geometric phase.
In our experimental implementation, we minimize $L$ under the constraint of a given pulse shape, which provides the curves with the least diabatic error for a given total length $z_{\textrm{\footnotesize{f}}} - z_{\textrm{\footnotesize{i}}}$ (see Supplementary Materials).

\section{Experiment}
In order to implement our findings, we employ a photonic platform manifested in the form of integrated coupled waveguides.
Using the analogy between the quantum evolution of a wavefunction and the propagation of an optical wavepacket in the paraxial approximation \cite{Longhi}, the quantum wells in our structure can be replaced by optical waveguides.
The temporal evolution of the light amplitudes in those waveguides is governed by Eq.~(\ref{eq:Hamiltonian}) with the sole difference that the time evolution is replaced by the evolution along the waveguides described by the spatial coordinate $z$ (see Fig.~2~A).
Our design protocol yields a spatial evolution of the intersite hoppings $Q$, $S$, $P$ with an example depicted in Fig.~2~B.
The hoppings $S$ and $P$ resemble the Stokes and pump pulses of Gaussian shape in the counterintuitive sequence known from STIRAP \cite{STIRAP}, with $Q$ as an additional constant coupling.
The evolution of the parameters in parameter space is chosen to form a closed-loop trajectory as shown in Fig.~2~C.
Therefore, this evolution results in a geometric phase.

\begin{figure}
 \center
 \includegraphics[width=8cm]{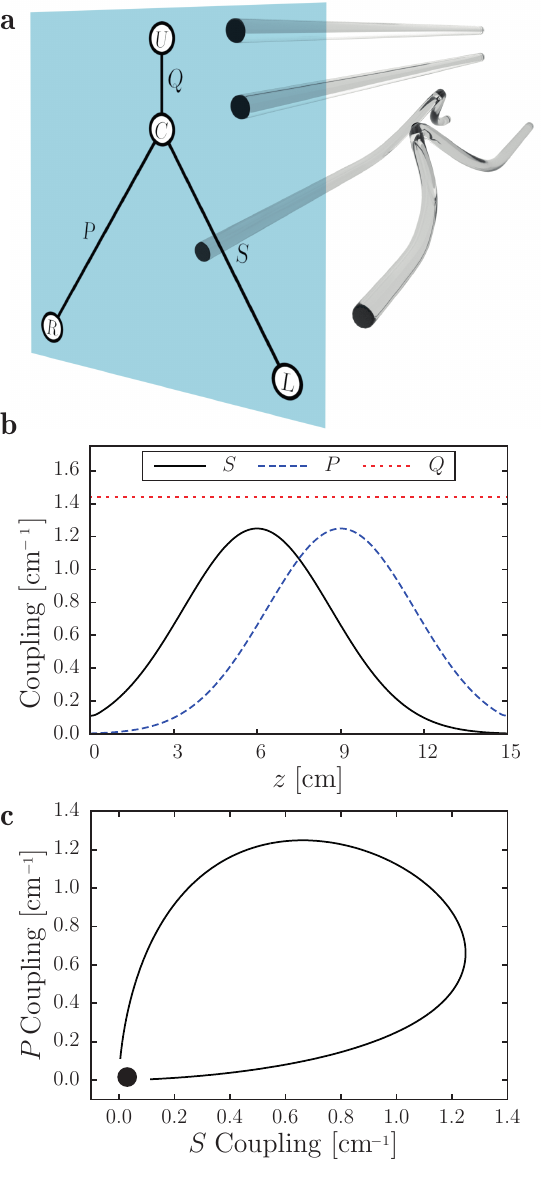}
 \caption{\textbf{a} 3D-rendering of the waveguides for one realization. \textbf{b} Coupling variation along propagating distance $z$. \textbf{c} Curve in the parameter manifold $\{S,P,Q\}$ ($Q=\textrm{const}$).}
\end{figure}

In the following, we will describe the measurement protocol for retrieving the Wilson loop. From the choice of the temporal evolution of the parameters $S$ and $P$, we have at the input facet $z_{\textrm{\footnotesize{i}}}$ and the output facet $z_{\textrm{\footnotesize{f}}}$ the relations $P(z_{\textrm{\footnotesize{i}}})/S(z_{\textrm{\footnotesize{i}}})\simeq 0\simeq S(z_{\textrm{\footnotesize{f}}})/P(z_{\textrm{\footnotesize{f}}})$.
Hence, the dark states at both facets simply become $|\mathcal{D}_1(z_{\textrm{\footnotesize{i}}})\rangle = -|w_{\textrm{\footnotesize{R}}}\rangle$, $|\mathcal{D}_2(z_{\textrm{\footnotesize{i}}})\rangle = |w_{\textrm{\footnotesize{L}}}\rangle$ and $|\mathcal{D}_1 (z_{\textrm{\footnotesize{f}}}) \rangle = |w_{\textrm{\footnotesize{L}}} \rangle$, $|\mathcal{D}_2 (z_{\textrm{\footnotesize{f}}}) \rangle = |w_{\textrm{\footnotesize{R}}} \rangle$.
As a consequence, launching light into the waveguides $L$ and $R$ excites only the dark states of the system.
Also, measuring the light intensity emanating from the waveguides $L$ and $R$ at the output facet provides the information about the population transfer between the dark states.
An initial superposition of the dark states evolves according to a unitary evolution $\mathbf{\mathcal{U}}$.
As we show in the Supplementary Materials, the elements of this unitary matrix can be expressed in terms of the amplitudes of the dark states at the input and output facet.
Moreover, the value of the Wilson loop is given by $W_{\mathcal{C}}=\mathrm{Tr}\left[\mathbf{\mathcal{U}}\right]$.
Measuring the field intensities yields the absolute values of the matrix elements $|\mathcal{U}_{ki}|$, and hence the absolute value $|W_{\mathcal{C}}|$ as shown in the Supplementary Materials.

For the fabrication of our samples, we use the femtosecond laser writing technique \cite{Szameit}.
Details of the fabrication are given in the Methods section.
We realize several structures with varying temporal profiles of the coupling parameters $S$ and $P$, resulting in different values of the Wilson loop.
An example of the evolution along the waveguides recorded by fluorescence microscopy (see Methods) is shown in Fig.~3~A.
Launching light into waveguide $L$ excites only the dark state $|\mathcal{D}_2 (z_{\textrm{\footnotesize{i}}}) \rangle$.
During the evolution, the light is coupled to the waveguides $R$ and $U$ without ever populating $C$ (see Fig.~3~B).
Hence, the evolution indeed remains in the dark subspace for all times as required for adiabaticity.

\begin{figure}
 \center
 \includegraphics[width=8cm]{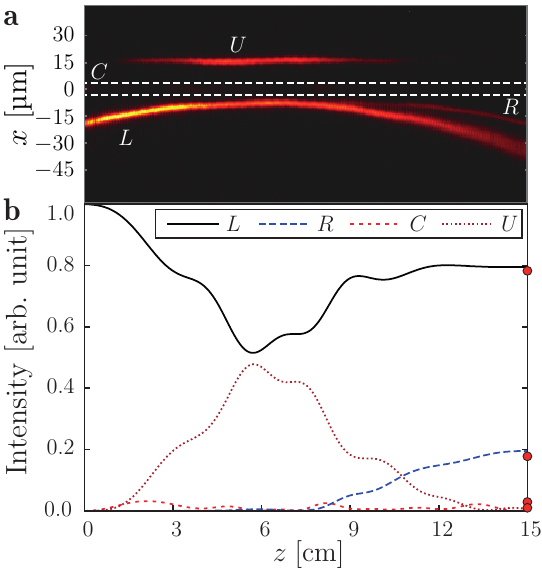}
 \caption{Intensity distribution in the four waveguides along the propagation distance.
 \textbf{a} Experimentally measured fluorescence signal which is proportional to the intensity in the waveguides.
 The waveguide L was excited, thus the second dark state.
 The central waveguide C is located between the dashed lines, highlighting the almost vanishing intensity.
 \textbf{b} Theoretically predicted intensity from the coupled-mode theory using Eq.~(\ref{eq:Hamiltonian}).
 Red dots at the end facet are the experimentally measured intensities (compare Fig.~4).}
\end{figure}

For retrieving the elements of $\mathbf{\mathcal{U}}$, we measure the intensities at the output facet.
A representative example is shown in Fig.~4~A. In our experiments, we realized Wilson loops by implementing three different sets of parameters (details of which are given in the Supplementary Materials).
The results are summarized in Tab.~\ref{tab:results}, where the theoretical predictions and the experimental results are shown to agree well.
In all three cases, the (absolute) value of the  Wilson loop is well above $0$, thus proving the non-Abelian character of the underlying contour.


\begin{table}
\caption{\label{tab:results} \textbf{Experimental results and  corresponding theoretical Wilson loops for three different coupling variations}.
For detailed description of the used parameters and pulse shapes see the Supplementary Materials.}
\centering
\begin{tabular}{cc}
\hline\hline $|W_{\mathcal{C}}^{\textrm{\footnotesize{theo}}}|$ & $|W_{\mathcal{C}}^{\textrm{\footnotesize{exp}}}|$ \\
 \hline
 0.88 & 0.87\\
 0.97 & 1.00\\
 1.07 & 1.13\\\hline\hline
\end{tabular}
\end{table}

In order to prove that waveguide $C$ is part of the full eigenspace, we specifically launched light into $C$ and excited states in the bright subspace that extend over all waveguides (see Fig.~4~B).
From our measurements, we find that at the output facet, all waveguides are bright such that one can conclude that, indeed, waveguide $C$ has to reside within the bright subspace.

\begin{figure}
 \center
 \includegraphics[width=8cm]{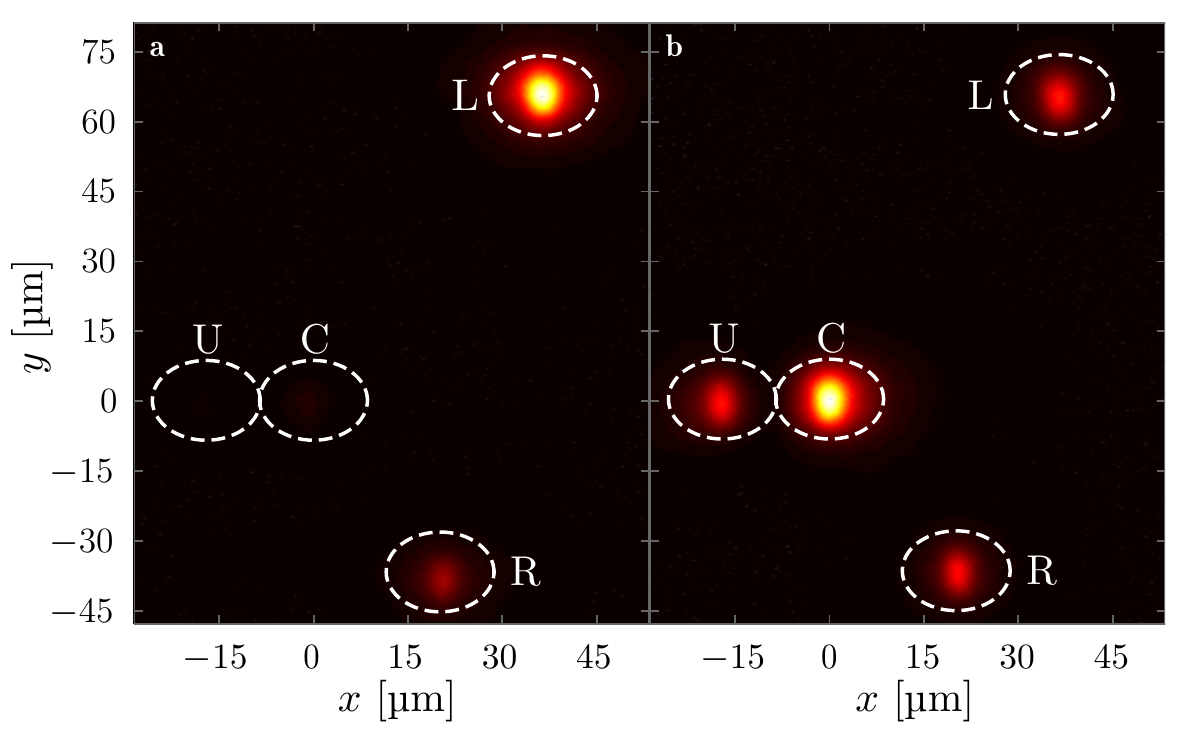}
 \caption{Measured intensities at the end facet. \textbf{a} Intensity distribution for dark-state excitation.
 This is for the same set of parameters as seen in Fig.~3.
 The result is clearly restricted to the dark subspace.
 \textbf{b} Intensity distribution for bright-state distribution.
 The waveguide C and U, and with that the bright states, are dominantly excited.}
\end{figure}

\section{Conclusion}
We employed evanescently coupled photonic waveguides to simulate the action of a non-Abelian gauge field on the dark subspace of the associated Hamiltonian.
The non-Abelian nature of the process was verified by measurement of the gauge invariant Wilson loop.
As the present implementation of the Wilson loop requires an adiabatic evolution within the dark subspace, the quantum metric is the appropriate tool to quantify the diabatic error.

Our results lay the foundations for the simulation of non-Abelian gauge fields using Abelian systems such as sound waves, matter waves or, in our case, light.
In particular, within this construction principle, the implementation of non-Abelian gauge fields that transform under SU($N$) could be realized with $N+2$ coupled sites containing an $N$-dimensional dark subspace.
Moreover, the use of geodesics of the quantum metric to quantify adiabaticity sheds new light on the optimization of all STIRAP-type processes.

The implementation of non-Abelian Abelian gauge fields prompts various important questions.
One of them concerns the simulation of lattice gauge field theories such as Yang--Mills theories where Wilson loops are the observable quantities.
In another context, using nonclassical light, our proposed setting is conducive to realize holonomic quantum operations as quantum logical gates can be defined as the action of non-Abelian geometric phases on the space of degenerate states, i.e. the dark subspace \cite{Pachos}.
Finally, the definition of a quantum metric induced by parametric changes of the waveguide couplings allows to study the evolution of a quantum system on curved manifolds.

\section{Acknowledgments}
The authors acknowledge funding from the Deutsche Forschungsgemeinschaft (grants SCHE 612/6-1, SZ 276/20-1, SZ 276/15-1, BL 574/13-1, SZ 276/9-2) and the Krupp von Bohlen and Halbach foundation.

\clearpage


\section*{Supplementary material}

# Supplementary Materials for

# Optimal design strategy for non-Abelian geometric phases using Abelian gauge fields based on quantum metric

Lucas Teuber*, Mark Kremer*, Alexander Szameit†, Stefan Scheel

Institut für Physik, Universität Rostock,
Albert-Einstein-Straße 23, D-18059 Rostock, Germany

† To whom correspondence should be addressed; E-mail: alexander.szameit@uni-rostock.de
* Both authors contributed equally

## Derivation of gauge field

A quantum state $|\psi\rangle$ initially prepared at $z = z_\mathrm{i}$ in the $N_\mathcal{D}$-dimensional dark subspace, i.e.

$$|\psi(z_\mathrm{i})\rangle = \sum_{j=1}^{N_\mathcal{D}} c_j |\mathcal{D}_j(z_\mathrm{i})\rangle \tag{1}$$

stays in this subspace if the evolution is perfectly adiabatic. At a later point $z > z_\mathrm{i}$ the state is thus given by applying to the initial state a unitary matrix $\mathcal{U} \in U(N_\mathcal{D})$,

$$|\psi(z)\rangle = \mathcal{U}(z, z_\mathrm{i})|\psi(z_\mathrm{i})\rangle = \sum_{i,j=1}^{N_\mathcal{D}} \mathcal{U}_{ij}\, c_j |\mathcal{D}_i(z)\rangle, \tag{2}$$

given by

$$\mathcal{U}(z, z_\mathrm{i}) = \sum_{i,j=1}^{N_\mathcal{D}} \mathcal{U}_{ij} |\mathcal{D}_i(z)\rangle\langle \mathcal{D}_j(z_\mathrm{i})|. \tag{3}$$

1

Hence, for initial states with $c_j = \delta_{jn}$, i.e. initially exciting $|\mathcal{D}_n(z_i)\rangle$, and measuring $|\mathcal{D}_m(z_f)\rangle$ yields the element $\mathcal{U}_{mn}$.

Inserting Eq. (2) into the Schrödinger or paraxial wave equation and acknowledging the fact that the state stays in the dark subspace with zero eigenvalue leads to

$$\partial_z|\psi(z)\rangle = \sum_{i,j=1}^{N_\mathcal{D}} c_j \left(\partial_z \mathcal{U}_{ij}|\mathcal{D}_i(z)\rangle + \mathcal{U}_{ij}\partial_z|\mathcal{D}_i(z)\rangle\right) = 0. \tag{4}$$

The second term on the right-most side is due to the apparent change of the dark states, i.e. the basis vectors of the dark subspace. When a state is thus infinitesimally transported along a curve it is not only the state that changes in its original basis but the basis itself. This is directly analogous to the transport of a vector in curved space where the additional derivative of the basis vectors gives the information of how to *connect* vectors along infinitesimal steps of the curve. In differential geometry this connection is described by the well-known Christoffel symbols, whereas in case of the dark state dynamics this information is encoded in the gauge field. Thus, Eq. (4) not only describes the adiabatic transport of the state $|\psi(z)\rangle$ in the dark subspace but can also be interpreted geometrically as a parallel transport, where the total change is zero.

The emergence of a gauge field from the change of basis occurring in Eq. (4) can be understood as follows. Since Eq. (4) has to hold for arbitrary input states, i.e. $\forall c_j$, we have

$$\sum_{i=1}^{N_\mathcal{D}} \partial_z \mathcal{U}_{ij}|\mathcal{D}_i(z)\rangle = -\sum_{i=1}^{N_\mathcal{D}} \mathcal{U}_{ij}\partial_z|\mathcal{D}_i(z)\rangle. \tag{5}$$

Applying $\langle \mathcal{D}_k(z)|$ from the left results in

$$\partial_z \mathcal{U}_{kj} = -\sum_{i=1}^{N_\mathcal{D}} A_{ki}\mathcal{U}_{ij}, \tag{6}$$

where

$$A_{ki} = \langle \mathcal{D}_k(z)|\partial_z|\mathcal{D}_i(z)\rangle. \tag{7}$$

As a non-Abelian gauge field the $N_\mathcal{D} \times N_\mathcal{D}$-matrix $\mathbf{A}$ has to fulfil the proper transformation

behaviour under change of basis

$$|\tilde{\mathcal{D}}_i\rangle = \sum_j U_{ij}|\mathcal{D}_j\rangle \tag{8}$$

with a unitary matrix $\mathbf{U}$. The new gauge field is then

$$\tilde{A}_{ki} = \sum_{mj} A_{mj} U^*_{km} U_{ij} + \sum_m U^*_{km} \partial_z U_{im}, \tag{9}$$

$$\tilde{\mathbf{A}} = \mathbf{U}\mathbf{A}\mathbf{U}^{-1} + (\partial_z \mathbf{U})\mathbf{U}^{-1}, \tag{10}$$

which is a properly transformed gauge field [3].

A general solution to Eq. (6) up to $z = z_\text{f}$ is the path-ordered integral

$$\mathcal{U}(z_\text{f}) = \mathcal{P}\exp\left(-\int_{z_\text{i}}^{z_\text{f}} \mathbf{A}\,\mathrm{d}z\right). \tag{11}$$

Since the system parameters depend on the propagation distance $z$ we have a curve $\mathcal{C} : (z_\text{i}, z_\text{f}) \to \mathcal{M}$ in the parameter manifold $\mathcal{M}$ with coordinates $\{x^\nu\}$. By using the total derivative, we find

$$(\mathbf{A}_\nu)_{ki} = \langle \mathcal{D}_k | \frac{\partial}{\partial x^\nu} | \mathcal{D}_i \rangle, \tag{12}$$

where $\mathbf{A}_\nu$ are matrix-valued the components of the gauge field expressed in the parameters of $\mathcal{M}$. Subsequently, for a closed curve $\mathcal{C}$ we get

$$\mathcal{U} = \mathcal{P}\exp\left(-\oint_\mathcal{C} \mathbf{A}_\nu \mathrm{d}x^\nu\right). \tag{13}$$

This can be characterized by the related gauge-invariant quantity, the so called Wilson loop [11]

$$W_\mathcal{C} = \mathrm{Tr}\,\mathcal{U}. \tag{14}$$

## Gauge field of tripod STIRAP

Choosing the coordinates $\{x^\nu\} = \{S, P, Q\}$ that parameterize the tripod STIRAP system, cf. Fig. 1 A, the dark states take the form

$$|\mathcal{D}_1\rangle = P/(\sqrt{S^2 + P^2})|w_\text{L}\rangle - S/(\sqrt{S^2 + P^2})|w_\text{R}\rangle, \tag{15}$$

3

$$|\mathcal{D}_2\rangle = SQ/(\sqrt{S^2+P^2+Q^2}\sqrt{S^2+P^2})|w_\mathrm{L}\rangle \tag{16}$$

$$+PQ/(\sqrt{S^2+P^2+Q^2}\sqrt{S^2+P^2})|w_\mathrm{R}\rangle - \sqrt{S^2+P^2}/(\sqrt{S^2+P^2+Q^2})|w_\mathrm{U}\rangle. \tag{17}$$

Calculating the $\mathbf{A}_\nu$ from Eq. (12) yields

$$\mathbf{A}_S = \mathrm{i}\, PQ/((S^2+P^2)\sqrt{S^2+P^2+Q^2})\,\sigma_y, \tag{18}$$

$$\mathbf{A}_P = -\mathrm{i}\, SQ/((S^2+P^2)\sqrt{S^2+P^2+Q^2})\,\sigma_y. \tag{19}$$

Since the gauge field is only proportional to the Pauli matrix $\sigma_y$ we solve the path-ordered Eq. (13) and give directly the result as [8]

$$\mathcal{U} = \begin{pmatrix} \cos\gamma & \sin\gamma \\ -\sin\gamma & \cos\gamma \end{pmatrix} \tag{20}$$

with

$$\gamma = \int_{z_\mathrm{i}}^{z_\mathrm{f}} (Q(S\partial_z P - P\partial_z S))/((S^2+P^2)\sqrt{S^2+P^2+Q^2})\mathrm{d}z \tag{21}$$

and

$$W_\mathcal{C} = 2\cos\gamma. \tag{22}$$

Three remarks are important at this point. First, in order to check the non-trivial nature of the implemented gauge field we could test if $|W_\mathcal{C}| < 2$, meaning the system cannot be described as a collection of Abelian subsystems. However, even though a process with $|W_\mathcal{C}| < 2$ is often referred to as non-Abelian [19] this criterion is only necessary but not sufficient for a "truly" non-Abelian gauge field[20]. Nevertheless, it can be regarded as non-Abelian geometric phase since it is a non-trivial collection of interacting Abelian subsystems. Therefore, we use the above criterion to characterize our implemented gauge fields.

Second, in the special case occurring in the experiments, the dark state designation to the waveguides L and R flip from $z_\mathrm{i}$ to $z_\mathrm{f}$, i.e. $|\mathcal{D}_1(z_\mathrm{i})\rangle = -|w_R\rangle$, $|\mathcal{D}_2(z_\mathrm{i})\rangle = |w_L\rangle$ and $|\mathcal{D}_1(z_\mathrm{f})\rangle = |w_L\rangle$, $|\mathcal{D}_2(z_\mathrm{f})\rangle = |w_R\rangle$. This results in a change of the above criterion to $|W_\mathcal{C}| > 0$. However,

4

this is only due to the flip in designation and could also be remedied by an additional index flip in the definition of $\mathcal{U}_{ij}$ from which we abstained. This has no repercussions on the applicability of the criterion or its statement.

Third, we mention that we measured the intensity at the end facets of the waveguide system and therefore we can only retrieve the absolute values $|\mathcal{U}_{ij}|$. However, since we deal only with real couplings $S$, $P$, and $Q$ no relative phases occur meaning the elements of $\mathcal{U}$ are also real and we can retrieve at least $|W_\mathcal{C}|$. An extension would be the inclusion of phase measurements by interference which can be done in the waveguide system with a few extra waveguides.

## Quantum Metric

Interpretation of the generalized STIRAP process as a dynamic under the influence of a gauge field is not the only way to look at it. In a completely equivalent way we interpret the evolution of a state as propagating in a curved (metric) space, i.e. our parameter manifold defined by the couplings $\{x^\nu\} = \{S, P, Q\}$. Traversing the curve $\mathcal{C}$ in this manifold is thus governed by a non-trivial metric tensor providing us with a measure of length along $\mathcal{C}$. A sensible definition [14, 15] of such a metric tensor is given by the infinitesimal change of the dark subspace projector $P(z) = \sum_i |\mathcal{D}_i(z)\rangle\langle\mathcal{D}_i(z)|$,

$$||\mathrm{d}P||^2 = \mathrm{Tr}\,(\mathrm{d}P\mathrm{d}P) = \mathrm{Tr}\,(\partial_\mu P \partial_\nu P)\,\mathrm{d}x^\mu\mathrm{d}x^\nu = g_{\mu\nu}\mathrm{d}x^\mu\mathrm{d}x^\nu, \qquad (23)$$

with the metric tensor being $g_{\mu\nu} = \mathrm{Tr}\,(\partial_\mu P \partial_\nu P)$. This metric tensor is the real part of the *quantum metric tensor* from the literature [13, 14, 15, 16].

Any process initiated by a parameter variation can be assigned a path length or action along the curve $\mathcal{C}$ in the parameter manifold which is defined via the metric tensor, i.e.

$$L = \int_{z_\mathrm{i}}^{z_\mathrm{f}} \sqrt{g_{\mu\nu}\partial_z x^\mu \partial_z x^\nu}\mathrm{d}z. \qquad (24)$$

Minimization with the variational principle leads to geodesics that define processes with least diabatic error for a given total length along $z$ [14]. If we have an actual experimental implementation with certain boundary conditions on curves that can be realized, we might not be able to obtain perfect geodesics. Nevertheless, $L$ is a possible measure for the quality of a certain curve/parameter variation and may be used for optimization, i.e. helps to find the curve with the least diabatic error/path length for given constraints (like pulse shape etc.). This idea was used to obtain optimized coupling variations when designing the waveguide system for different gauge fields or Wilson loops with details on the couplings discussed in the next section.

A few comments are in order at this point. The introduced gauge field connection and metric tensor are similar in that both are ultimately linked to the non-vanishing derivative of the dark states. The first by the projection of the derivative on the dark subspace, cf. Eq. (12), and the latter by the derivative of the dark subspace projector itself, cf. Eq. (23). However, the metric tensor **g** has its own connection or Christoffel symbol and is not linked to the gauge field in the sense of a metric connection in differential geometry. This is rooted in the fact that the gauge field is not completely analogous to the Christoffel symbol since the former connects dark states and the latter tangential vectors along a curve $\mathcal{C}$. Hence, the elements $(\mathbf{A}_\nu)_{ki}$ incorporate the parameter manifold index $\nu$ and dark subspace indices $k$, $i$. As a result, $(\mathbf{A}_\nu)_{ki}$ cannot be expressed in terms of derivatives of a metric like the Christoffel symbols can. Indeed, as a metric connection the gauge field is only associated with the trivial metric $\delta_{ij}$ of the dark subspace. For further discussion of the metric tensor and its intricate relation to the gauge field connection see [15, 21].

For completeness we also give the metric tensor that describes the tripod STIRAP process with the three real couplings $S$, $P$, and $Q$. In this case we find

$$\mathbf{g} = 2\,\text{diag}\left(1, \cos^2(\phi)\right). \tag{25}$$

Note that the metric is two-dimensional since the mixing angles $\theta$ and $\phi$ are the relevant parameters here. This is because the normalization of the dark states makes one parameter degree of $S$, $P$, and $Q$ obsolete.

## Details on Couplings

For an experimental measurement of the process described by the holonomy $\mathcal{U}$ we have to fix the curve $\mathcal{C}$ in parameter space. However, in the theory outlined above we saw that in order for the evolution being restricted to the dark subspace and thus $\mathcal{U}$ being valid we require adiabaticity. This requirement is now opposed by experimental boundary conditions which might limit the number of possible curves and by that the number of unitaries $\mathcal{U}$ that can be realized. For example a *hard* boundary condition is the physical length of the glass chips in which the photonic waveguides are laser-written. We used $15$ cm glass chips which therefore set the maximal total propagation length and with that the slowest possible process. Using the quantum metric, or the path length $L$ defined by it, we can find those coupling variations that comply with the experimental constraints and result in the most adiabatic processes ensuring the validity of $\mathcal{U}$.

For the temporal variation of the couplings we choose consecutive Gaussian pulses for $S$ and $P$ (counter-intuitive pulse sequence) and hold $Q = \Omega_Q$ constant. The Gaussian pulse sequences are parameterized by

$$S(z) = \Omega \exp\left(-(z - \bar{z} + \tau)^2 / T^2\right), \tag{26}$$

$$P(z) = \Omega \exp\left(-(z - \bar{z} - \tau)^2 / T^2\right), \tag{27}$$

where $\bar{z}$ is half the total propagation length, $\Omega$ the amplitude, $T$ the width parameter and $\tau$ the separation of the two Gaussian pulses from the center at $\bar{z}$ (given by length of the glass chip). This parameterization allows for simple fabrication of the waveguides and limit the number of parameters in the optimization. In addition, if we restrict the parameters $\Omega$, $\tau$, and $T$ so that we

Table 2: **Coupling parameters and results**. Pulse parameters for the experimental realization of different pulse shapes/Wilson loop values.

| $\Omega_Q$ (cm$^{-1}$) | $\Omega$ (cm$^{-1}$) | $T$ (cm$^{-1}$) | $\tau$ (cm$^{-1}$) | $|W_\mathcal{C}^{\text{theo}}|$ | $|W_\mathcal{C}^{\text{exp}}|$ |
|---|---|---|---|---|---|
| 1.42 | 1.23 | 3 | 1.5 | 0.88 | 0.87 |
| 1.53 | 1.46 | 3 | 1.5 | 0.97 | 1.00 |
| 1.60 | 1.8  | 3 | 1.5 | 1.07 | 1.13 |

achieve $P(z_\text{i})/S(z_\text{i}) \approx 0$ and $S(z_\text{f})/P(z_\text{f}) \approx 0$ the dark states at these points become simply

$$|\mathcal{D}_1(z_\text{i})\rangle = -|w_R\rangle, \qquad |\mathcal{D}_1(z_\text{i})\rangle = |w_L\rangle, \tag{28}$$

$$|\mathcal{D}_2(z_\text{i})\rangle = |w_L\rangle, \qquad |\mathcal{D}_2(z_\text{f})\rangle = |w_R\rangle. \tag{29}$$

Additionally, we restrict the coupling amplitudes $\Omega_Q$ and $\Omega$ to be smaller than $2\text{cm}^{-1}$ to ensure that the tight-binding approximation is still valid.

The optimization procedure is then the following. For sets of the four parameters $\Omega_Q$, $\Omega$, $T$, and $\tau$ that comply with the above conditions the path length $L$ is calculated. Each set also resulted in a different theoretical value for the Wilson loop. Therefore we binned all sets according to their Wilson loops and chose those that minimize $L$ for a given loop. Since for increasing values of the Wilson loop (mind the flip as discussed in main text the maximal coupling strength also sharply increases to ensure adiabaticity, which is why we could only collect sets up to a Wilson loop of $\approx 1.2$. However, for loops below that we could collect parameter sets that minimize $L$ and result in different theoretical Wilson loop values.

In the experiments we implemented three of those sets with the optimized parameters seen in Tab. 2. Also listed are the theoretical and experimental values of the Wilson loops.



\end{document}